\documentclass{aastex}
\usepackage{url}
\usepackage{graphicx}
\usepackage{epstopdf}
\usepackage{amsmath}
\usepackage{amssymb}
\RequirePackage{color}

\newcommand{\emaila}{mlazar@tp4.rub.de}

\def\be{\begin{equation}}
\def\ee{\end{equation}}
\def\bdm{\begin{displaymath}}
\def\edm{\end{displaymath}}

\begin{document}

\title{Instability constraints for the electron temperature anisotropy in the slow solar wind. 
Thermal core vs. suprathermal halo.}

\author{M.~Lazar,\altaffilmark{1,2} S.M.~Shaaban,\altaffilmark{2,3} V.~Pierrard,\altaffilmark{4,5} 
H.~Fichtner,\altaffilmark{1} S.~Poedts\altaffilmark{2}} 

\altaffiltext{1}{Institut f\"ur Theoretische Physik, Lehrstuhl IV: Weltraum- und Astrophysik, 
Ruhr-Universit\"at Bochum, D-44780 Bochum, Germany, \emaila } 

\altaffiltext{2}{Centre for Mathematical Plasma Astrophysics, Celestijnenlaan 200B, 3001 Leuven, Belgium}

\altaffiltext{3}{Theoretical Physics Research Group, Physics Dept., Faculty of Science, 
Mansoura University, 35516, Egypt}

\altaffiltext{4}{Royal Belgian Institute for Space Aeronomy, av. Circulaire 3, B-1180 Brussels, Belgium}

\altaffiltext{5}{Center for Space Radiations, Universit\'e Catholique de Louvain,
Place Louis Pasteur 3 bte L4.03.08, 1348 Louvain-La-Neuve, Belgium}


\begin{abstract}

This letter presents the results of an advanced parametrization of the solar wind electron 
temperature anisotropy and the instabilities resulting from the interplay of the (bi-)Maxwellian 
core and (bi-)Kappa halo populations in the slow solar wind. A large set of observational data 
(from the Ulysses, Helios and Cluster missions) is used to parametrize these components and establish 
their correlations. The instabilities are significantly stimulated in the presence of suprathermals, 
and the instability thresholds shape the limits of the temperature anisotropy for both the core and 
halo populations re-stating the incontestable role that the selfgenerated instabilities can play in 
constraining the electron anisotropy. These results confirm a particular implication of the 
suprathermal electrons which are less dense but hotter than thermal electrons.
\end{abstract}



\section{Observational motivation}

The velocity distributions of electrons in the solar wind are well represented by 
a combination of a Maxwellian core at low-energies and Kappa power-law at higher energies, 
suggesting the existence of two distinct (but inter-correlated) populations, namely
a thermal core and a suprathermal Kappa-distributed halo \citep{ma05,st08, pi16}.
An additional strahl component drifting along the magnetic field in antisunward direction
can also be observed, but mainly during the energetic events (e.g., fast winds
and coronal mass ejections) and mainly at small (less than 1~AU) heliocentric distances \citep{ma05}.
Otherwise the implication of strahls is minimal, and the electron 
dynamics is controlled by the interplay of the core and halo populations.


The solar wind electrons exhibit a temperature anisotropy with respect to the (local) 
magnetic field direction, but the anisotropy does not increase indefinitely
as would result from different concurrent processes, such as the fluid-like expansion, 
magnetic compression, or dissipation (at lower scales) via wave-particle resonances.
At large radial distances particle-particle collisions are not efficient and the selfgenerated instabilities 
may act constraining large deviations from isotropy. The observations have already confirmed this scenario
but only for the core components of the solar wind electrons \citep{st08} and protons \citep{he06}. 
To be precise, the instability thresholds predicted by the linear theory for these bi-Maxwellian populations 
have been found to approach well enough the limits of temperature anisotropy observed in the solar wind. 

Here we analyze the implications of the suprathermal electrons which are ubiquitous in the solar 
wind. Being more tenuous and hotter than the core, suprathermal populations may be sources of 
free energy for the selfgenerated instabilities \citep{vi10, wi13}, which are expected to grow faster and 
exhibit lower threshold conditions in the presence of suprathermals \citep{sh16, la17}. 
If we refer to the solar wind electrons, the core and halo components cannot exist independently, and 
their interplay can be inferred from the correlations of different quantities, e.g., temperature anisotropy
and (parallel) plasma beta, which are the principal parameters conditioning the instability thresholds. 
Distinctive descriptions revealing correlations of these two populations are not abundant in the literature.
Early studies were limited in characterizing the halo as a remainder after measuring the core 
(bi-Maxwellian fitting or numerical integration) and extracting from the total distribution \citep{mc92, ph95}, 
or by fitting both components with standard bi-Maxwellians \citep{pi87,ma97,ma00}. A more 
accurate description has been provided later with a dual fitting model combining a bi-Maxwellian core 
and a bi-Kappa halo \citep{ma05,st08,pi16}. Emphasis has been given to the evolution of these two 
electron components as a function of heliocentric distance and latitude, 
mainly from an attempt to understand the origin of suprathermal electrons as well as their implication in the 
regulation of the energy budget and key processes (like plasma heating and particle acceleration) in the solar 
wind. The relative halo to core density is roughly constant over heliocentric distance with the halo 
representing 4~\% of the total electron density \citep{mc92}, but  out of the ecliptic it shows an 
increasing tendency to reach values of 10 to 30~\% of the total density \citep{ma05} and a steep 
radial gradient \citep{ma00}. For the halo (subscript $h$) 
to core (subscript $c$) temperature contrast \cite{ma00} have found a modest variation with the solar wind bulk
speed, i.e., mean values $T_h/T_c \simeq 13.57$ in the slow wind ($V_{SW} < 600$ km/s), and $T_h/T_c 
\simeq 23.38$ in the fast wind ($V_{SW} > 600$ km/s), although their bi-Maxwellian representation for 
the halo component cannot reproduce details of the distribution which may be important in the analysis 
of the temperature anisotropy instabilities. These authors have also studied the so-called 
suprathermal strength $S \equiv n_h T_h/n_c T_c$ (the ratio of the halo to core kinetic pressures), which is 
actually giving the correlation of the halo and core plasma beta parameters, i.e., $S=\beta_h/\beta_c$ 
(where $\beta = 8 \pi n k_B T/B_0^2$ is the ratio of the kinetic and magnetic field pressures), and found
average values higher in the fast ($S\simeq 0.79$) than the slow ($S \simeq 0.39$) wind.

The temperature anisotropies of the electron core and halo populations have been measured by \cite{st08}
for more than 120.000 events detected in the ecliptic by three spacecraft (Helios, Cluster and Ulysses)
and covering the radial distances from the Sun from 0.3 up to 4 AU. Parallel and perpendicular temperatures
($T_{\parallel,\perp}$, with respect to the local magnetic field) are determined from fitting 
the observed velocity distribution (after instrumental corrections) with a gyrotropic dual model that combines 
a bi-Maxwellian core with a bi-Kappa halo. Limits of the temperature anisotropy ($A = T_\perp/T_\parallel$)
observed in the solar wind have been compared with the instability thresholds predicted by the linear
kinetic theory for a bi-Maxwellian distributed plasma, i.e., the whistler instability driven by an excess
of perpendicular temperature $T_\perp > T_\parallel$, and the firehose instability driven by $T_\parallel >
T_\perp$. Good agreements between these instability thresholds and the limits of the temperature anisotropy 
are found only for the electron core population in the slow winds (i.e., with a bulk speed $V < 500$~km/s.
The halo component shows significant departures from the anisotropy thresholds especially for the 
firehose instability. Even stronger differences are obtained for both the core and halo populations
in the fast wind ($V > 600$~km/s), but our analysis in the present paper focuses on the slow wind 
conditions. The disagreement obtained for the halo component must have an immediate explanation (also 
suggested by the authors) in the facts that (\textit{i}) the theoretical model used in the linear prediction of the 
instabilities (bi-Maxwellian) is different than that used to reproduce the observed halo data (bi-Kappa),
and (\textit{ii}) theoretical predictions for the halo component neglect completely the possible effect of the 
core. In this letter we present the results of an advanced stability analysis, intended to decode the
interplay of the electron core and halo populations from a realistic parametrization in accord with 
the fitting model. The core-halo correlations between physical quantities, e.g., relative densities, 
anisotropies and (parallel) plasma betas are identified from the observations in section~\ref{sec2}. 
We then derive the instability thresholds for the whistler and firehose instabilities and 
compare with the limits of temperature anisotropy observed in the solar wind (section~\ref{sec3}). 
The results are discussed in section~\ref{sec4}.

\section{Core-halo correlations} \label{sec2}

\cite{pi16} have used the same set of electron data invoked by \cite{st08} to investigate the solar wind 
electron temperature and its anisotropy making a distinction between the core and halo 
components, and also suggesting a series of correlations which we invoke in the present 
analysis. Thus, the halo temperature shows a clear variation with the power-index $\kappa$, which 
quantifies the presence of suprathermals in a bi-Kappa distribution, and \cite{la17}
have found for this dependence
\be T_h = T_M \kappa / (\kappa - 1.5), \label{e1} \ee
where $T_M$ is a (Maxwellian) limit value. This law is in perfect agreement with the radial evolution
of the halo temperature that increases while the power-index $\kappa$ increases.

\begin{figure}[t]
\centering 
\includegraphics[width=145mm]{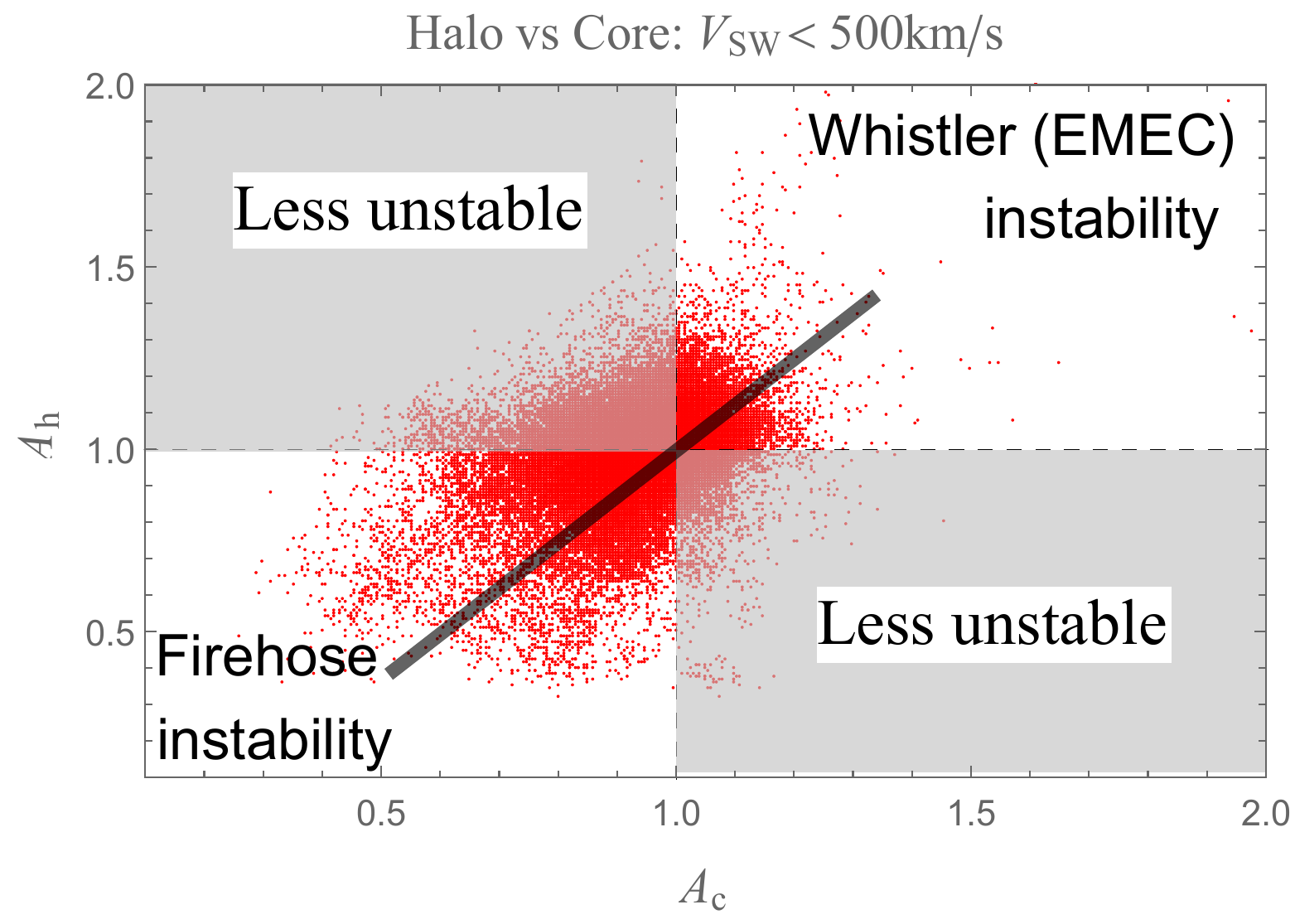} 
\caption{Temperature anisotropies $A_h$ vs. $A_c$ in slow winds ($V_{\rm SW} < 500$ km$/$s).} \label{f1}
\end{figure}

Plots correlating the temperature anisotropies of the core and halo components are provided
in Fig.~5 from \cite{pi16}, where the data points show a prevailing disposition to align along
a linear regression $A_h = A_c$, suggesting directly correlated anisotropies. Here in Fig.~\ref{f1} 
we rebuild such a plot of $A_h$ vs. $A_c$, but with the full set of data. Linear regression $A_h = 
A_c$ is shown with a solid line, and the isotropic conditions $A_h = 1$ or $A_c = 1$ are drawn with dashed lines.
Dominant appear to be the states with $A_c < 1$, and among these most abundant are the symmetric states
with both $A_h < 1$ and $A_c < 1$, which are the most unstable against the excitation of the firehose
instability (FHI). The opposite quadrant contains states with both $A_h > 1$ and $A_c > 1$, which 
are the most unstable to the excitation of the electromagnetic electron cyclotron (EMEC) instability
also known as the whistler instability (WI). In the other two quadrants, the core and halo exhibit
opposite (or anticorrelated) anisotropies, i.e., $A_h > 1$ and $A_c < 1$ (top), or $A_h < 1$ and 
$A_c > 1$ (bottom). These plasma states are less unstable (a detailed study of the interplay of the 
core and halo instabilities will be presented elsewhere), and the resulting instabilities will not
evolve fast enough to be efficient in constraining the temperature anisotropy of the solar wind electrons.
Instead, the WI is highly stimulated when both $A_{c,h} > 1$, and the FHI when both $A_{c,h} < 1$. 

\begin{figure}[t]
\centering 
\includegraphics[width=75mm]{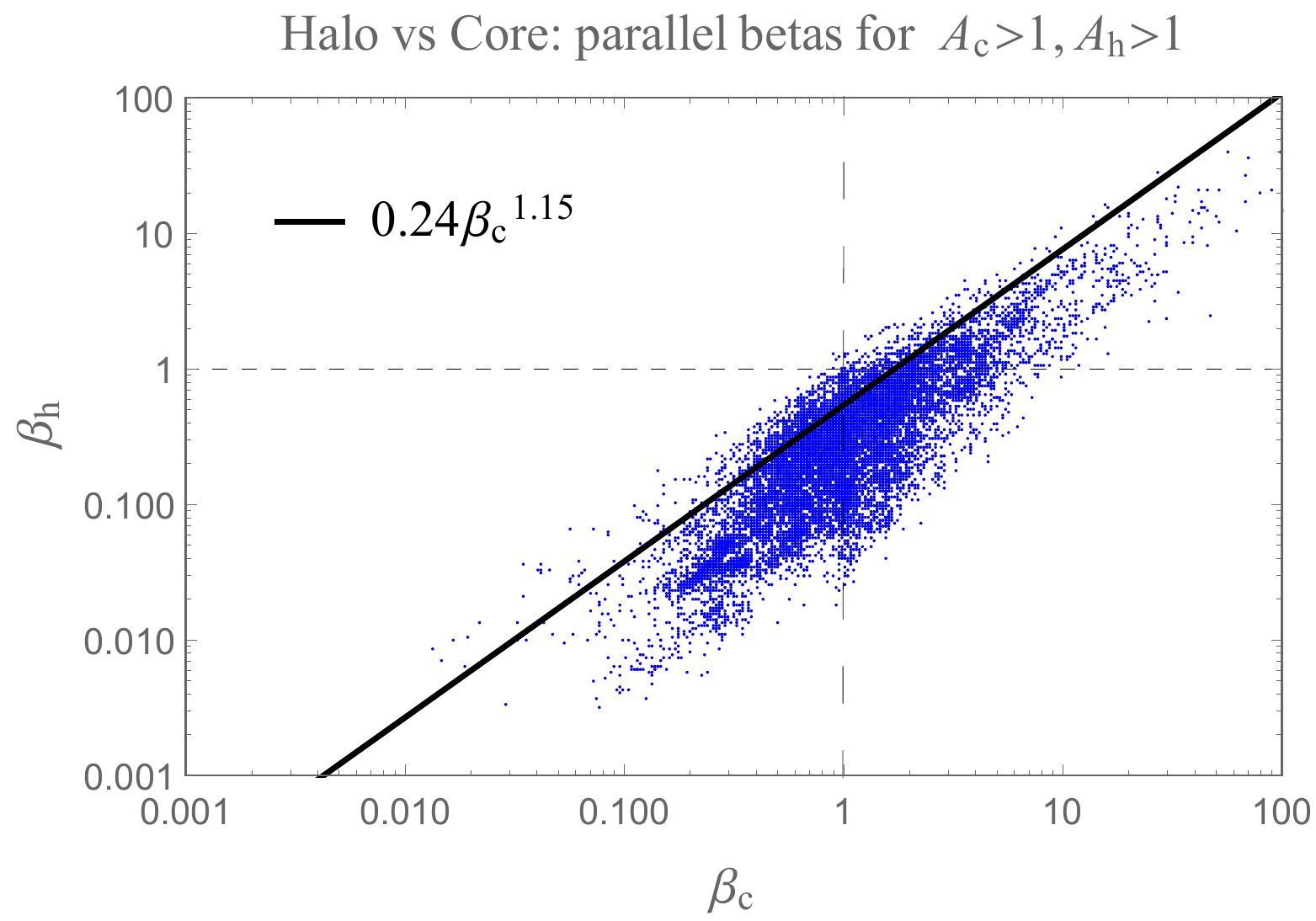}  \includegraphics[width=75mm]{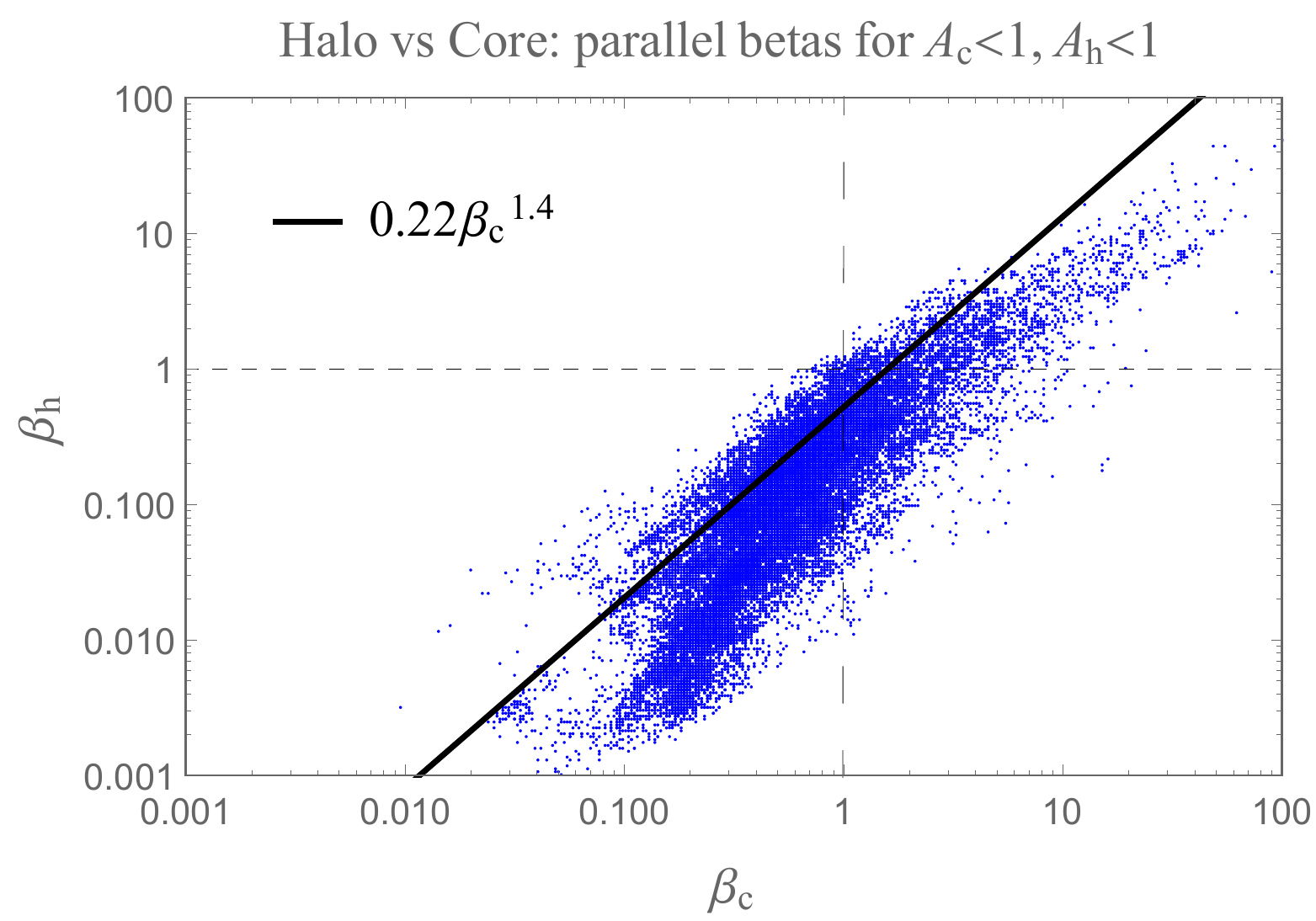}
\caption{Parallel plasma betas $\beta_{h,\parallel}$ vs. $\beta_{c,\parallel}$ in slow winds 
($V_{\rm SW} < 500$ km$/$s), for $A_{c,h} > 1$ (top) and $A_{c,h} < 1$ (bottom).} \label{f2}
\end{figure}

In Figure~\ref{f2} we display $\beta_{h,\parallel}$ vs. $\beta_{c,\parallel}$ for each of these two cases
of interest here, and the best linear fitting (log-log scales) plotted over with solid 
lines provides $\beta_{h,\parallel} = 0.24 \beta_{c,\parallel}^{1.15}$ for $A_{c,h} > 1$ (top panel),
and $\beta_{h,\parallel} = 0.22 \beta_{c,\parallel}^{1.4}$ for $A_{c,h} < 1$ (bottom panel).
Other key parameters in the dispersion/stability analysis are the relative density numbers of electrons
in the core $\eta_c = n_c/n_0$ and halo $\eta_h = n_h/n_0$ components ($n_0 = n_{\rm total}$ is the total 
density number), which are evaluated using the observational data in Table~1 from \cite{pi16}. We 
find the following mean values $\eta_c = 0.966$ for the core, and $\eta_h = 0.034$ for the halo, 
implying for the halo-core number density contrast $\eta= n_h/n_c = 0.035$, almost the same with
that obtained by \cite{mc92} in similar conditions in the ecliptic.

%

\section{Instability thresholds vs. observations} \label{sec3}

The stability analysis may invoke the electromagnetic cyclotron modes, which are
particularly important in establishing resonant interactions with magnetized plasma particles,
and constraining their temperature anisotropy. The fastest growing mode driven by the 
electrons with $A >1$ is the parallel whistler instability (also known as the electromagnetic 
electron cyclotron mode) with growth rates higher than the other oblique modes 
including the propagating whistlers \citep{ke66} and the non-propagating mirror instability 
\citep{ga06} which can be ignored \citep{st08}. High-frequency whistlers are right-handed 
polarized modes which do not interact with protons and are described by the dispersion relation
\begin{align} \tilde{k}^2= & \dfrac{1}{\eta}\left[A_{c}-1+\frac{A_{c}\left(\tilde{\omega}-1 \right)+1}
{\tilde{k} \sqrt{\eta \beta_{c,\parallel}}}Z_M\left(\frac{\tilde{\omega}-1}{\tilde{k} 
\sqrt{\eta \beta_{c,\parallel}}}\right) \right] \nonumber \\
& +A_{h}-1+\frac{A_{h}\left(\tilde{\omega}-1 \right)+1}{\tilde{k} \sqrt{\beta_{h,\parallel}^M}} 
Z_\kappa\left(\frac{\tilde{\omega}-1}{\tilde{k} \sqrt{\beta_{h,\parallel}^M}}\right) 
\end{align}
where $\tilde{\omega}=\omega /\Omega= \tilde{\omega}_r +i \tilde{\gamma}$ (including the wave-frequency
$\tilde{\omega}_r$ and the growth-rate $\tilde{\gamma}$) and wave-number 
$\tilde{k}=~kc/\omega_{e,h}$ are normalized by, respectively, the electron gyro-frequency 
$\Omega = |\Omega_e|$ and the halo plasma frequency $\omega_{e,h}$, and $Z_M$ and $Z_\kappa$ are 
the plasma dispersion functions for the Maxwellian and Kappa distributed plasmas (see for instance
the definitions in \cite{sh16}). $\beta_{h,\parallel}^M$ corresponds to the Maxwellian (limit) 
temperature in Eq.~\ref{e1}, $ \beta_{h,\parallel} = \beta_{h,\parallel}^M \kappa/(\kappa-1.5)$.
In parallel direction the low-frequency electron FHI is a left-handed polarized mode described by 
\begin{align} 
\tilde{k}^2=& \eta_c~\mu \left[A_{c}-1+\frac{A_{c}\left(\tilde{\omega}+\mu \right)-\mu}
{\tilde{k} \sqrt{\mu\beta_{c,\parallel}/\eta_c}}Z_M\left(\frac{\tilde{\omega}+\mu}{\tilde{k} \sqrt{\mu \beta_{c,\parallel}/\eta_c}}\right)\right] \nonumber \\
& +\eta_h \mu \left[A_{h}-1+\frac{A_{h}\left(\tilde{\omega}+\mu \right)-\mu}{\tilde{k} 
\sqrt{\mu \beta_{h,\parallel}^M/\eta_h}} Z_\kappa\left(\frac{\tilde{\omega}+\mu}{\tilde{k} \sqrt{\mu \beta_{h,\parallel}^M/\eta_h}}\right)\right] \nonumber \\
&+ \dfrac{\tilde{\omega}}{\tilde{k} \beta_{p,\parallel}}Z_M\left(\frac{\tilde{\omega}-1}
{\tilde{k} \sqrt{\beta_{p,\parallel}}}\right), 
\end{align}
where the frequency $\tilde{\omega}=\omega /\Omega _{p}= \tilde{\omega}_r +i \tilde{\gamma}$ and wave-number 
$\tilde{k}=~kc/\omega_{p}$ are normalized by, respectively, the proton gyro-frequency 
$\Omega_p$ and proton plasma frequency $\omega_{p}$. The influence of protons  ($\mu=m_p/m_e=1836$) 
is minimized assuming isotropic and of finite temperature, with $\beta_{p,\parallel} = 
\beta_{c,\parallel}/2$.

\begin{figure}[t]
\centering 
\includegraphics[width=100mm]{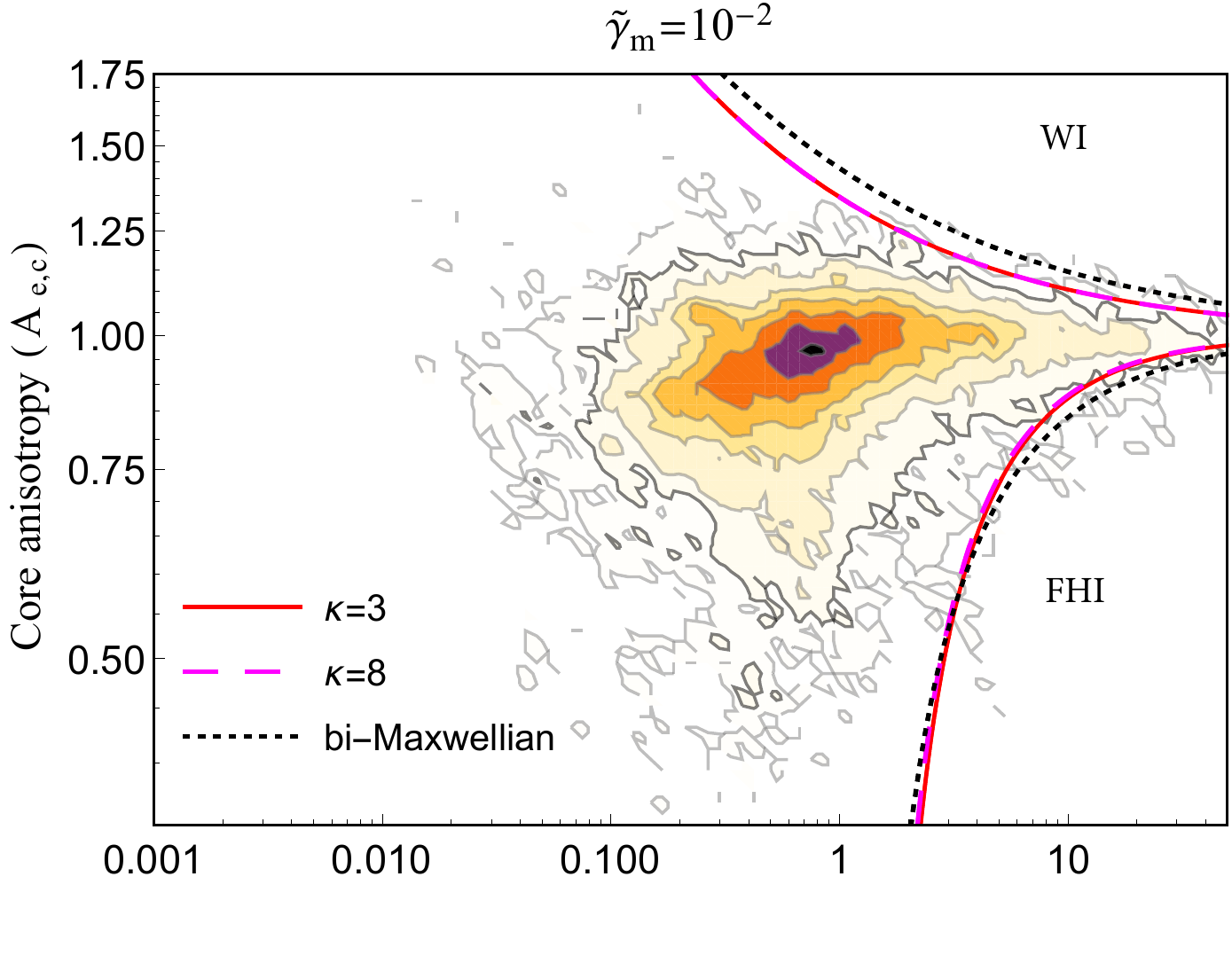} \\ \includegraphics[width=100mm]{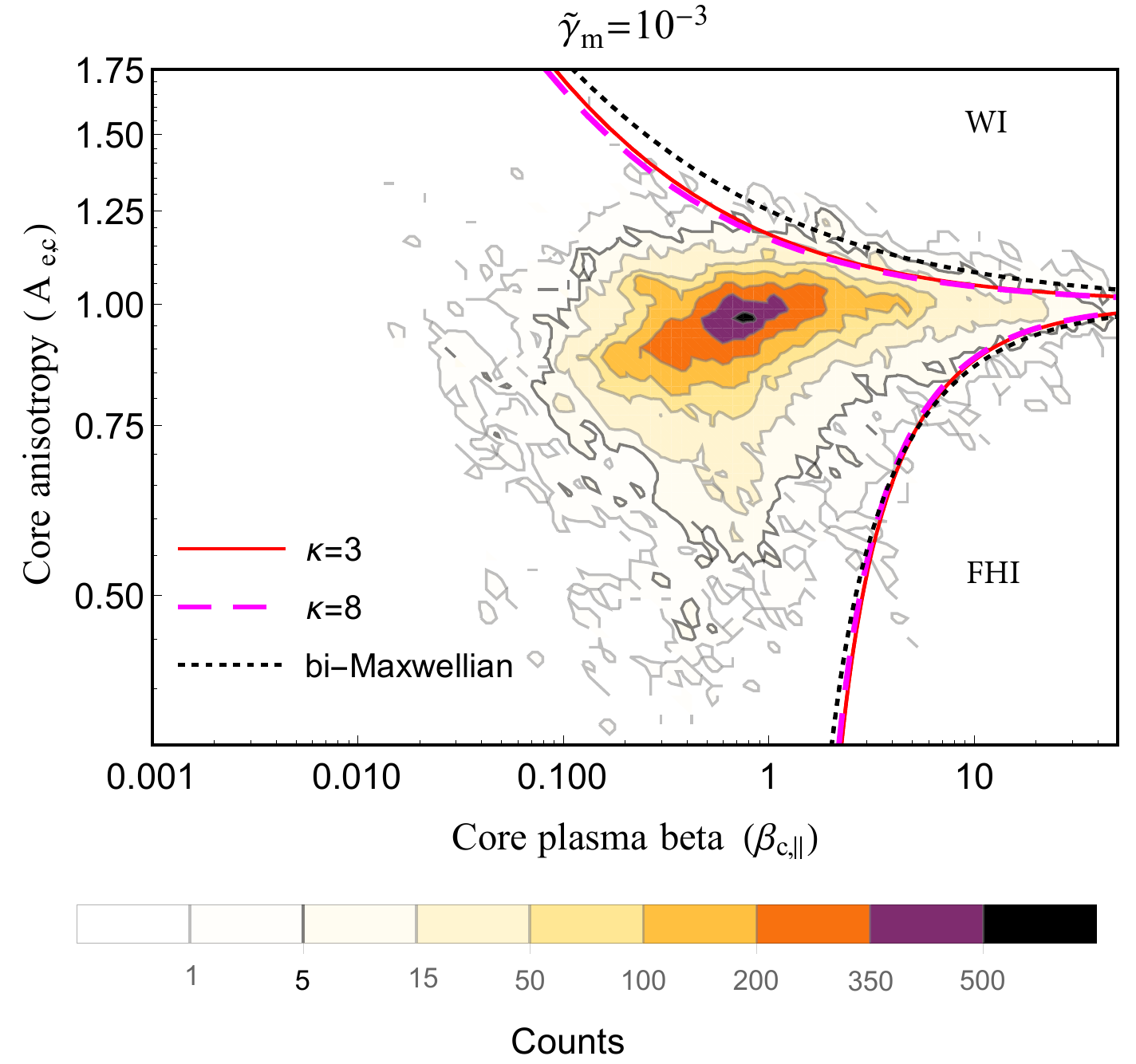}
\caption{Comparison of the instability (WI and FHI) thresholds for $\tilde{\gamma}_m = 10^{-3}$ 
and 10$^{-2}$, with the core temperature anisotropy displayed using a histogram data (counts in color 
bar).} \label{f3}
\end{figure}

\begin{figure}[t]
\centering 
\includegraphics[width=100mm]{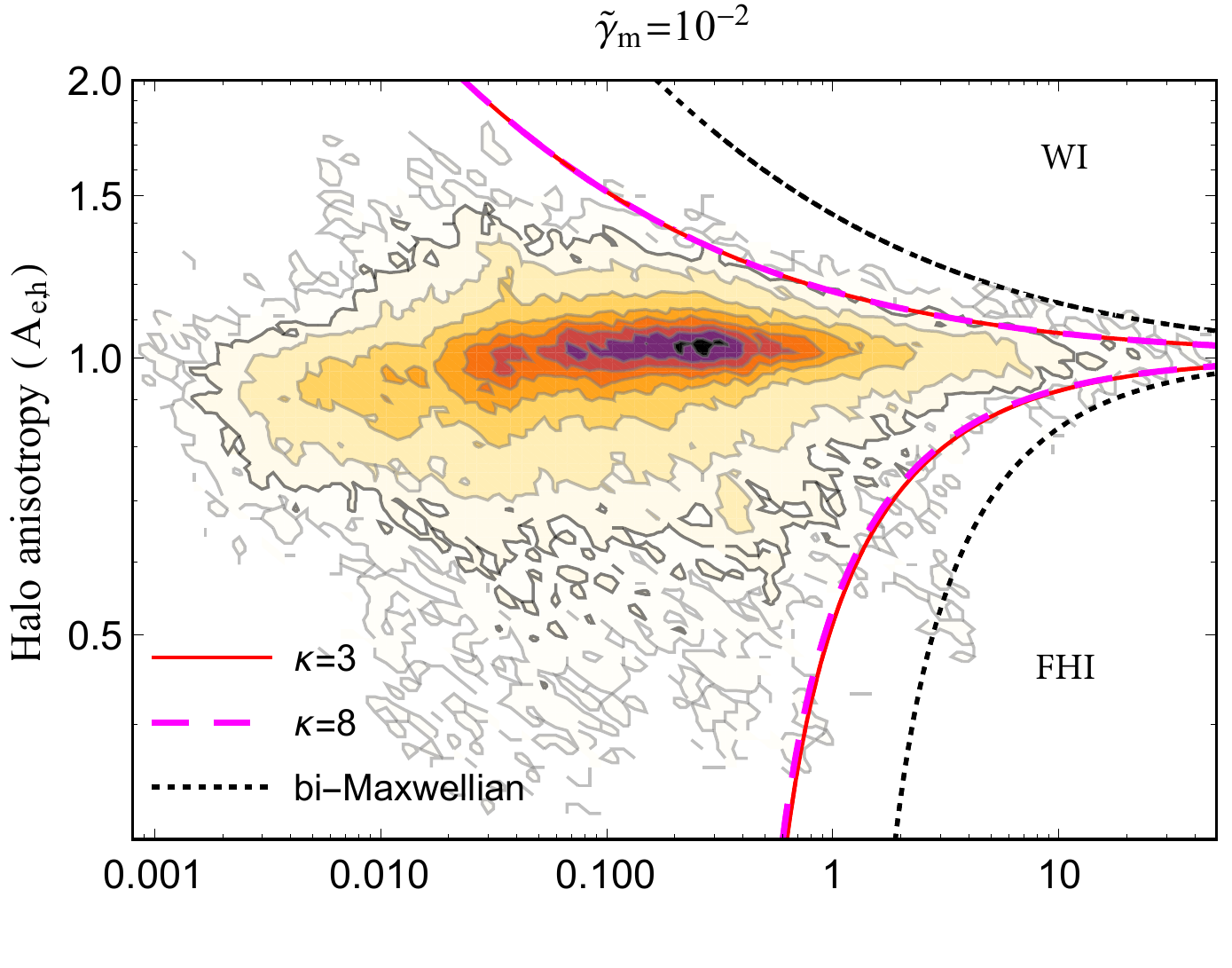} \\ \includegraphics[width=100mm]{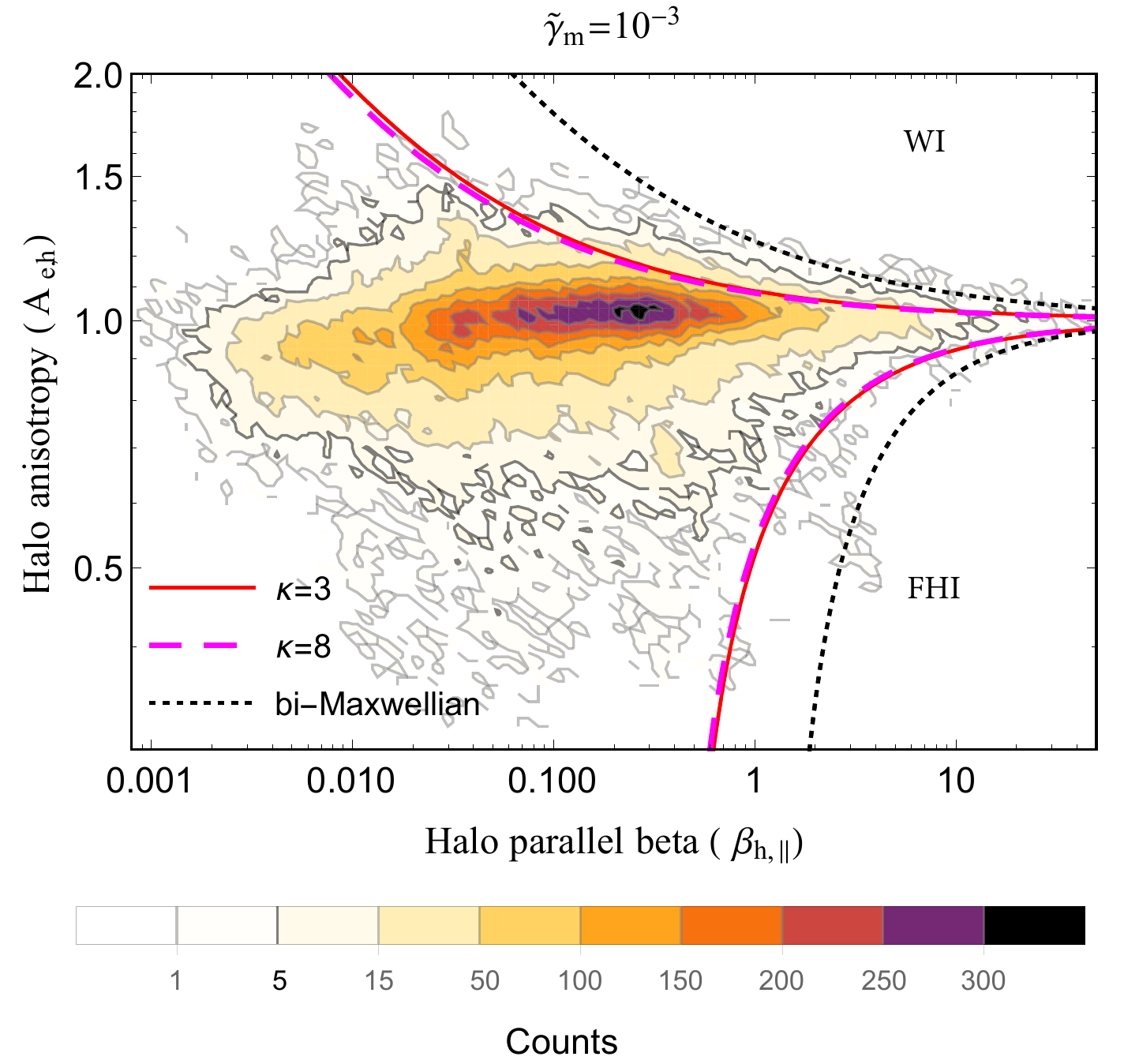}
\caption{Comparison of the instability (WI and FHI) thresholds for $\tilde{\gamma}_m = 10^{-3}$ 
and 10$^{-2}$, with the halo temperature anisotropy displayed using a histogram data (counts in color 
bar).} \label{f4}
\end{figure}

We use these dispersion relations and the parametrization established in previous section to derive the 
temperature anisotropy thresholds as curves of constant (maximum) growth rates $\tilde{\gamma}_m = 10^{-3}$, 
10$^{-2}$. The anisotropy curves are described as inverse correlation laws of the parallel plasma 
beta, i.e., $A = 1 + a/\beta_\parallel^b$, and values obtained for the fitting parameters $a$ and $b$ are 
tabulated in Tables~\ref{t1} and \ref{t2}. The instability conditions are determined for two
values of the power-index $\kappa = 3$ and 8, representative for sufficiently many events 
with, respectively, the lowest and highest $\kappa-$values from our data set. In Figures~\ref{f3} and 
\ref{f4} thresholds of WI and FHI are contrasted with the temperature anisotropy measured in the solar 
wind, for the thermal (Maxwellian) core (Figure~\ref{f3}) and for the suprathermal (Kappa) halo 
(Figure~\ref{f4}). The distribution of observational data providing $A$ vs. $\beta_\parallel$ is 
shown with histogram contours to count the events.

\begin{table}[t]
\centering \caption{Parameters $a$, $b$ for the core anisotropy thresholds.  } \label{t1}
\begin{tabular}{c c c c c c c c c c}
\hline
Wave & \multicolumn{4}{c}{WI} & & \multicolumn{4}{c}{FHI} \\ 
\hline
$\tilde{\gamma}_m$& \multicolumn{2}{c}{$10^{-3}$} &  \multicolumn{2}{c}{$10^{-2}$} & & 
\multicolumn{2}{c}{$10^{-3}$} & \multicolumn{2}{c}{$10^{-2}$} \\ 
\hline
                    & $a$  & $b$     & $a$     &$b$    &  & $-a$        &$b$     &$-a$  & $b$ \\ 
$\kappa=3$ &0.18&  0.59 &0.34  &0.53  &  &1.60 &1.10  &1.60 &1.10 \\ 
$\kappa=8$ &0.16&  0.60 &0.34  &0.53  &  &1.55 &1.11  &1.55 & 1.10\\
bi-Max.               &0.25&  0.50 & 0.43 &0.47 &   &1.76 &1.04  &1.70 & 0.99\\
\hline  
\end{tabular}
\end{table}

\section{Discussion and conclusion} \label{sec4}

The in-situ measurements of the velocity distributions of electrons in space plasmas (e.g., the 
solar wind and planetary magnetospheres) indicate the existence of two distinct but 
intercorrelated components, namely, a thermal core at low energies and a less dense but more 
energetic suprathermal halo. We have derived the temperature anisotropy thresholds of the 
whistler and firehose instabilities from the interplay of the core and halo populations, which 
are parametrized on the basis of a large set of observational data from Ulysses, Helios and 
Cluster missions. The correlations established between the core and halo parameters which 
determine the growth of instability (i.e., temperature anisotropy, parallel plasma beta and 
relative number density) enabled us to derive the instability thresholds function of either 
the core or halo parameters. Comparing with the observations, we find that the instability 
thresholds (solid and dashed lines) shape the limits of the temperature anisotropy for both 
the core (Figures~\ref{f3}) and halo (Figures~\ref{f4}) populations. By these results linear 
stability predicts for the first time that the selfgenerated instabilities may regulate the 
anisotropic halo component. The instability constraints appear to be markedly strengthened 
by the cumulative effect of these two components as the new refined thresholds are lower 
than those predicted for a bi-Maxwellian core (dotted lines) by neglecting the influence of 
the suprathermal (bi-Kappa) halo. We are aware of the fact that for bi-Maxwellian 
electrons with $A < 1$ the nonpropagating FHI develops obliquely to the magnetic field and 
faster (with higher growth-rates and lower thresholds) than the parallel FHI \citep{li00,ca08}. 
Indicated by the observations, the new Maxwellian-Kappa model invoked here is more complex 
and make the analysis of the oblique modes less straightforward. However, an 
extended investigation on the full wave-vector spectrum is strongly motivated by the
the results reported in the present paper and will make the object of our next studies.

\begin{table}[t]
\centering \caption{Parameters $a$, $b$ for the halo anisotropy thresholds. } \label{t2}
\begin{tabular}{c c c c c c c c c c}
\hline
Wave & \multicolumn{4}{c}{WI} & & \multicolumn{4}{c}{FHI} \\ 
\hline
$\tilde{\gamma}_m$& \multicolumn{2}{c}{$10^{-3}$} &  \multicolumn{2}{c}{$10^{-2}$} & & 
\multicolumn{2}{c}{$10^{-3}$} & \multicolumn{2}{c}{$10^{-2}$} \\ 
\hline
                    & $a$  & $b$     & $a$     &$b$    &  & $-a$        &$b$     &$-a$  & $b$ \\ 
$\kappa=3$ &0.09&  0.513 & 0.17 &0.45  &  &0.48 &0.79  &0.49 &0.78 \\ 
$\kappa=8$ &0.08&  0.516 & 0.17 &0.45  &  &0.47 &0.79  &0.47 & 0.79\\
\hline  
\end{tabular}
\end{table}
%


%
%

%
%
%
%
%
%

\begin{acknowledgments}
M.L. acknowledges support from the Katholieke Universiteit Leuven, Ruhr-University Bochum 
and Alexander von Humboldt Foundation. These results were obtained in the framework of the 
projects SCHL~201/35-1 (DFG--German Research Foundation), G0A2316N (FWO--Vlaanderen), 
GOA/2015-014 (KU Leuven), and C~90347 (ESA Prodex). This research has been funded by the 
Interuniversity Attraction Poles Programme initiated by the Belgian Science Policy Office (IAP P7/08 CHARM). 
S.M.S. would like to thank the Egyptian Ministry of Higher Education for supporting 
his research activities.
Thanks are due to \v{S}. \v{S}tver\'{a}k for providing the observational data.
\end{acknowledgments}




%

%
%

\end{document}